\documentclass[reprint,aip,jcp,amsmath,amssymb,floatfix]{revtex4-1}

\usepackage{graphicx}

\newcommand{\bignone}{}
\newcommand{\tmmathbf}[1]{\ensuremath{\boldsymbol{#1}}}
\newcommand{\tmop}[1]{\ensuremath{\operatorname{#1}}}

\begin{document}

\title{Effect of molecular rotation on enantioseparation}
\author{Andreas Jacob}
\author{Klaus Hornberger}
\affiliation{University of Duisburg--Essen, Faculty of Physics,
Lotharstra{\ss}e 1-21, 47048 Duisburg, Germany}

\date{22 March 2012}

\begin{abstract}
Recently, several laser schemes have been proposed to separate racemic mixtures of enantiomers by splitting a molecular beam into subbeams consisting of molecules of definite chirality [Y.~Li, C.~Bruder, and C.~P. Sun, Phys.\ Rev.\ Lett.\ {\bf{99}}, 130403 (2007); X.~Li and M.~Shapiro, J.\ Chem.\ Phys.\ {\bf{132}}, 194315 (2010)]. These ideas rely on laser-induced effective gauge potentials in an adiabatic basis which lead to a chirality dependent force on the center-of-mass. However, the effect of molecular rotation has been neglected in these studies. Accounting for the full molecular quantum state we find that the potentials from the adiabatic dressed state approach cannot be recovered once the molecular orientation dynamics is included, even in the rotational ground state. This affects substantially the ability to perform enantioseparation in the above mentioned setups.
\end{abstract}

\maketitle

\section{Introduction}                                                                                                                                                                                                                                                                                                                                                                                                                                                                                                                                                                                                                                                                                                                                                                                                                                                                                                                                                  
Many molecules can exist in different, non-superimposable configurations, e.g.\ as left-handed $(L)$ or right-handed $(R)$ enantiomers \cite{woolley76,walker79,barron82,barron86,quack89,cina94,harris94,maierle98}. Quantum mechanically, the appearance of such chiral molecules may seem paradoxical from the point of view of first principles \cite{hund27}, but it can be explained by a dynamical stabilisation effect due to environmental decoherence, as effected by gas collisions \cite{harris81,harris82,trost09}. The purification of a racemic mixture is an important task in chemistry \cite{bodenhoefer97,mckendry98,zepik02}, due to the different biological and chemical properties of $L$ and $R$ enantiomers. This can be achieved by interconversion  \cite{shapiro91,salam98,fujimura99,shapiro00,thanopulos03} or, as we study here, by separating the left-handed molecules from their right-handed counterparts. 

When a near-resonant laser field is applied to a chiral molecule the sign of the complex Rabi frequency may depend on the chiral state \cite{kral01}. In Refs.~\onlinecite{li07,li10b} a closed loop scheme was suggested to make use of this phase difference. 
It yields a chirality dependent effective potential for the molecules in the laser field, which would allow one to separate enantiomers by splitting a racemic molecular beam into subbeams of definite chirality. However, the molecular orientation has been neglected in these proposals. 
In this article we study the effects of including the rotational state. This is shown to affect  substantially the ability to perform enantioseparation in the above mentioned proposals.

The structure of the article is as follows. In Sec.~\ref{sec:signsensitivity} we discuss  the topologies of transitions required for enantioseparation. The existence of a closed loop with at least one driven dipole transition \cite{kral01} is a crucial ingredient to the separation schemes \cite{li07,li10b}. In Sec.~\ref{sec:std} we briefly review how effective gauge potentials for the molecular motion are obtained by treating the molecular dynamics in the laser field in an adiabatic basis.  To incorporate rotation, we then  introduce in Sec.~\ref{sec:withrot} the corresponding wave functions and Hamiltonian and evaluate the selection rules. In Sec.~\ref{sec:numerics} we show that the adiabatic potentials cannot be recovered once molecular rotations are included, even in the rotational ground state. 

\section{Chirality dependent potentials \label{sec:signsensitivity}}
\begin{figure}[t,b]
  \begin{center}   
    \includegraphics[width=0.47\textwidth]{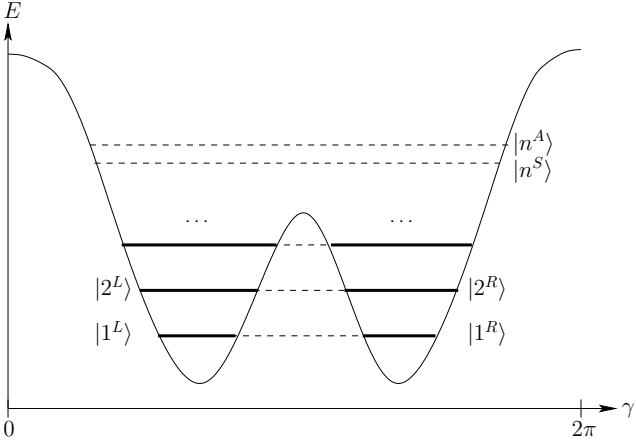}
  \end{center}
  \caption{Double well potential for the chiral configuration coordinate, such as the dihedral angle of $\rm D_2\rm S_2$.\label{fig:doublewell} The chiral molecular states, which are localized in one of the wells, are described by superpositions of the symmetric and the antisymmetric eigenstates of the Hamiltonian. }
\end{figure}

The laser-induced separation of stable chiral molecules relies on the fact that the complex Rabi frequencies describing the electric-dipole transition between two chiral states differ in sign for enantiomers. The left-handed and the right-handed molecular state, $| \Psi^{+} \rangle$ and $| \Psi^{-} \rangle$, can be described by a superposition of a symmetric and an antisymmetric eigenstate of the parity-invariant Hamiltonian \cite{shapirobook}, see Fig.~\ref{fig:doublewell},
\begin{eqnarray}
| \Psi^{\pm} \rangle & = & s |S \rangle \pm a |A \rangle .   
\end{eqnarray}
If an enantiomer is exposed to an electric laser field $\boldsymbol E$, the strength of the electric-dipole transition between chiral molecular states is described by the Rabi frequency
\begin{eqnarray}
  \Omega_{fi} & = & \langle \Psi_f^{\pm} | \hat{\boldsymbol\mu} | \Psi_i^{\pm} \rangle \cdot \boldsymbol{E} \, .
\end{eqnarray}
Since the diagonal element vanishes for parity eigenstates, we have
\begin{eqnarray}
    \Omega_{fi} & = & \pm [ s^{\ast}_f a_i  \langle S_f | \hat{\boldsymbol \mu }  |A_i \rangle + a_f^{\ast} s_i \langle A_f | \hat{\boldsymbol \mu} |S_i  \rangle ]  \cdot\boldsymbol{E} \, . 
\end{eqnarray}
Hence, two enantiomers see the same electric field, but the chiral sign difference is passed on to the Rabi frequency.

\begin{figure}[t,b,p]
  \begin{center}
    \includegraphics[width=0.47\textwidth]{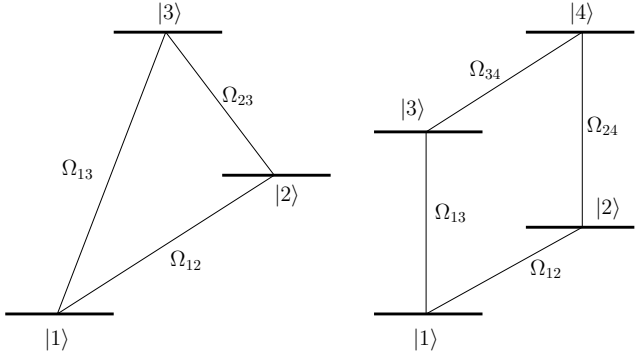}
 \end{center}
 \caption{Simple closed loop structures in the absence of rotational sublevels. Left: 3-level system connected by dipole transitions forming a loop. Right: simple 4-loop connecting four states. The dipole transition strength is given by the complex Rabi frequencies $\Omega_{fi}$. \label{fig:34loopsimple}}
\end{figure}

\begin{figure}[t,b,p]
  \begin{center}
    \includegraphics[width=0.4\textwidth]{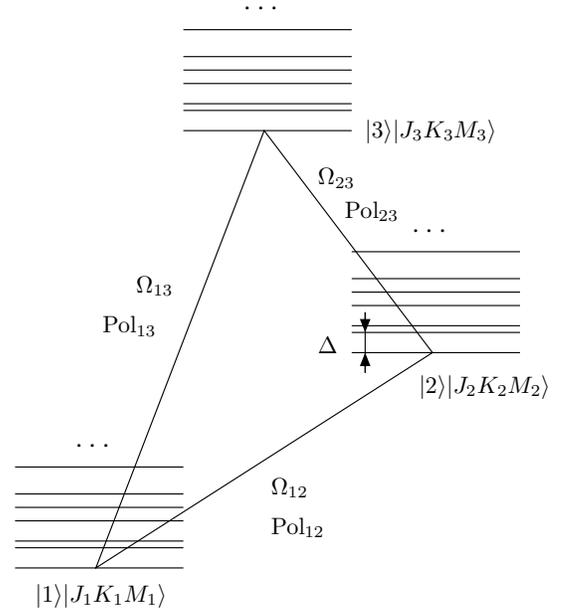}
  \end{center}
  \caption{Level scheme for a 3-loop setup with rotational substates\label{fig:3looprot}. The diagonal lines indicate the three lasers connecting the vibrational levels, with the vertical components describing the laser frequency. The endpoints do not necessarily connect an allowed transition, because the laser can also drive nearby transitions with detunings $\Delta$.}
\end{figure}

In the following we discuss what topological structures of the dipole transitions are required in order to ensure that the spectrum of the effective Hamiltonian is sensitive to a sign change of the Rabi frequency. The spectrum of the Hamiltonian then will give rise to the scalar potential $V$ which might be used to separate the enantiomers as discussed in Sec.~\ref{sec:std}. 

We consider a system of $N$ levels connected by laser-induced electric-dipole transitions. Starting with $N = 3$, the internal  Hamiltonian (consisting of bare states and the interaction) for a three-level system on resonance can in the interaction picture  be written as 
\begin{eqnarray}
  H_{\tmop{int}} & = & \left(\begin{array}{ccc} 
    0 & \Omega_{12}^{\ast} & \Omega_{13}^{\ast}\\
    \Omega_{12}^{} & 0 & \Omega_{23}^{\ast}\\
    \Omega_{13}^{} & \Omega_{23}^{} & 0
  \end{array}\right)  \, . \label{eq:3ham}
\end{eqnarray}
In this closed 3-loop setup, sketched in Fig.~\ref{fig:34loopsimple}, the eigenvalues of the 
internal Hamiltonian $H_{\tmop{int}}$ change only after a sign change of an {\em odd} number of Rabi transitions. This is seen by looking at the characteristic equation for a closed loop with three levels,
\[  - \varepsilon^3 + \varepsilon ( \left| \Omega_{12} \right|^2 +  \left| \Omega_{23} \right|^2 + \left| \Omega_{13} \right|^2) + 2 \tmop{Re}  [\Omega_{12} \Omega_{23} \Omega_{13}^{\ast}] \,   =  0 \, ,    \nonumber
\]
where $\varepsilon$ are the eigenvalues. The sign dependence is given by the last term. Similarly, for a 4-loop the characteristic equation
\begin{eqnarray}
 & &  \varepsilon^4 - \varepsilon^2 ( \left| \Omega_{34} \right|^2 + \left| \Omega_{24}
             \right|^2 + \left| \Omega_{12} \right|^2 + \left| \Omega_{13} \right|^2)   \label{eq:4characteristic} \\
  & &  \tmop{Re} [\Omega_{12}^{\ast} \Omega_{24}^{\ast} \Omega_{13}
             \Omega_{34}] + \left| \Omega_{12} \right|^2 \left| \Omega_{34} \right|^2 +
              \left| \Omega_{13} \right|^2 \left| \Omega_{24} \right|^2  =  0\nonumber
 \end{eqnarray}
shows that the eigenvalues vary only under a sign change of an odd number of Rabi frequencies $\Omega_{f i}$, because of the term $ \tmop{Re} [\Omega_{12}^{\ast} \Omega_{24}^{\ast} \Omega_{13} \Omega_{34}]$. One can verify also for 5-loops, and 6-loops that only an odd number of sign changes yields chiral control. For a general statement one should consider the Hamiltonian (\ref{eq:3ham}) as an adjacency matrix \cite{chen97}. A topological classification of possible terms appearing in a characteristic polynomial has to our knowledge only been carried out recently in \textcite{do2010}. A closer look at the characteristic equation of a system consisting of four levels with arbitrary transitions shows that only closed loop structures yield sign dependent energies. Possible double loops or subloops do not change this statement. However, it is impossible to embed the 4-loop shown in Fig.~\ref{fig:34loopsimple} with an odd number of sign changing transitions into the molecular level structure. For example, if we consider four chiral states in a 4-loop we will have four sign dependent transitions, which will not change the eigenvalues obtained from Eq.~(\ref{eq:4characteristic}). As one can check explicitly, this results hold even if we include parity eigenstates in the loop or any $n$-loop with even $n$.

The phase of the Rabi frequency $\Omega$ is irrelevant for most quantum optics effects, such as level shifts. The reason is that in systems with tree-like transition structures the phase of a `leg' can be removed by a gauge transformation, and has hence no physical meaning. If the link-structure contains a closed loop, on the other hand, the total phase associated to a loop can matter. In the next section we review how this setup can lead to chirality dependent dynamics.

\section{Scheme without rotational states \label{sec:std} }
Let us consider the simplest closed-loop system consisting of three molecular levels connected by three driven dipole transitions as in \textcite{li10b}. If one disregards the orientation degree of freedom, the Hamiltonian of the molecule in the laser field is
\begin{eqnarray}
  \hat{H}_{\tmop{tot}}^{(\chi)} & = & \hat{H}_{\tmop{CM}} + \hat{H}_{\tmop{int}}^{\left( \chi \right)}  + \hat{V} \, , 
\end{eqnarray}
where $\hat{H}_{\tmop{CM}}$ is the kinetic energy of the center-of-mass motion, $\hat{H}_{\tmop{int}}$ is the internal Hamiltonian including the molecule-laser interaction and the vibrational levels, $\hat{V}$ is a possible trapping potential, and $ \chi  \in \left\{ L, R \right\}$ denotes the chirality. For a fixed center-of-mass position $\tmmathbf{r}$ the internal Hamiltonian $\hat{H}_{\tmop{int}} ( \tmmathbf{r})$, which includes the space dependent interaction, can be diagonalized. This yields a set of dressed states $\left| \chi_n ( \tmmathbf{r}) \right\rangle$ with eigenvalues $\varepsilon_n ( \tmmathbf{r})$, where $n = 1, 2, 3$ since we focus on three levels. The full quantum state of the molecule describing internal and motional degrees of freedom can then be expanded in terms of the dressed states according to
\begin{eqnarray}
  | \Psi^{\left( \chi \right)} ( \tmmathbf{r})  \rangle & = & \sum_{n = 1}^3 \psi_n^{\left(
  \chi \right)} ( \tmmathbf{r}) | \chi^{\left( \chi \right)}_n ( \tmmathbf{r})
  \rangle . 
\end{eqnarray}
These dressed states obey an effective Schr\"odinger equation for $\boldsymbol{\psi}=\left(\psi_1,\psi_2,\psi_3 \right) $ \cite{wilczek84,ruseckas95,dalibard11}
\begin{eqnarray}
  \mathrm{i} \hbar \frac{\partial}{\partial t} \boldsymbol{\psi}^{\left( \chi \right)} & = &  \left[ \frac{1}{2 m} (- \mathrm{i} \hbar \tmmathbf{\nabla}  -  \tmmathbf{A^{}}^{\left( \chi \right)})^2 + V^{\left( \chi \right)} \right]  \boldsymbol{\psi}^{\left( \chi \right)}, \label{eq:effschr}
\end{eqnarray}
where the effective  potentials are given by $3 \times 3$ potentials
\begin{eqnarray}
  \tmmathbf{A}_{n m}^{\left( \chi \right)} & = & \mathrm{i} \hbar \langle
  \chi^{\left( \chi \right)}_n ( \tmmathbf{r}) | \tmmathbf{\nabla} 
  \chi^{\left( \chi \right)}_m ( \tmmathbf{r}) \rangle \\
  V_{nm}^{\left( \chi \right)} & = & \varepsilon^{\left( \chi \right)}_n (
  \tmmathbf{r}) \hspace{0.25em} \delta_{nm} + \langle \chi^{\left( \chi
  \right)}_n ( \tmmathbf{r}) |V (\tmmathbf{r}) | \chi^{(\chi)}_m ( \tmmathbf{r}) \rangle
  . 
\end{eqnarray}
The off-diagonal elements of the potentials can be neglected if they are much smaller than the eigenvalue differences. The equation then depends only on
\begin{eqnarray}
  \tmmathbf{A}^{\left( \chi \right)}_n & = & \mathrm{i} \hbar \langle
  \chi_n^{\left( \chi \right)} ( \tmmathbf{r}) | \tmmathbf{\nabla} 
  \chi_n^{\left( \chi \right)} ( \tmmathbf{r}) \rangle  \label{eq:vektorpot}\\
  V^{\left( \chi \right)}_n & = & \varepsilon^{\left( \chi \right)}_n (
  \tmmathbf{r}) \hspace{0.25em} + \langle \chi_n^{\left( \chi \right)} (
  \tmmathbf{r}) |V (\tmmathbf{r}) | \chi_n^{\left( \chi \right)} ( \tmmathbf{r}) \rangle
  .  \label{eq:skalarpot}
\end{eqnarray}
This way three decoupled equations of motion are obtained, each describing  a dressed state. To simplify the notation we drop the chiral index $\chi $ in the following.

The explicit forms of the potentials are given in Ref.~\onlinecite{li10b} for a setup of three resonant Gaussian beams. In this case the effect of the vector potentials $\tmmathbf{A}_n$ is smaller than the effect of the scalar potential $V_n$ that depends linearly on $\Omega$. The three scalar potentials $V_n$ for the three dressed states can have different properties. They may trap the molecule or accelerate it, depending on the dressed state and the chirality of the molecule. Based on this, it was suggest to use three levels of the $v_5$ vibrational mode of $\rm D_2 \rm S_2 $ in the electronic ground state to achieve a spatial separation \cite{li10b}. However, the real molecular state features also a rotational subspace which must be considered and will be taken into account in the next section.

\section{Accounting for the orientational state \label{sec:withrot}}
To allow for the molecular rotation dynamics one must include the rotational energy in the Hamiltonian and adapt the molecule-laser interaction, which depends on the molecular orientation.
Using the helicity basis, the laser interaction with the electric-dipole in the laboratory is given by $\hat{H}_{\tmop{IA}} = \tmmathbf{\hat{\mu}} \cdot \tmmathbf{E}= \sum_{\sigma \in \left\{ \pm 1, 0 \right\}}  \hat{\mu}^S_{\sigma} E^S_{\sigma}$. The components of the electric dipole in the space-fixed frame $(S)$ are obtained by a rotation from the molecular frame $(M)$:
\begin{eqnarray}
  &  & \hat{\mu}_{\sigma}^S \hspace{0.6em} = \hspace{0.6em} \sum_{\sigma' \in \left\{  \pm 1, 0 \right\}} D^{1 \ast}_{\sigma \sigma'} \left( \hat{\alpha}, \hat{\beta}, \hat{\gamma} \right) \hat{\mu}_{\sigma'}^M . 
\end{eqnarray}
Here $D$ is the rotation matrix \cite{zare88} and $\alpha, \beta, \gamma$ are the Euler angles, determining the orientation of the space-fixed $(S)$ relative to the molecule-fixed $(M)$ coordinate system. This yields the interaction Hamiltonian
\begin{eqnarray}
\hat{H}_{\tmop{IA}} & =&
\sum_{\sigma,\sigma' \in \left\{ \pm 1, 0 \right\} }
D^{1 \ast}_{\sigma \sigma'} \left(  \hat{\alpha}, \hat{\beta}, \hat{\gamma} \right) \hat{\mu}^M_{\sigma'}  E^S_{\sigma},   \label{eq:hintrot}
\end{eqnarray}
with $\sigma'$ indicating the spherical components of the dipole $\hat{\tmmathbf{\mu}}^M$ in the molecular frame, and $\sigma$ the helicity components of the electric field. 

Since we are interested in the Rabi frequencies $\Omega_{fi} = \langle \Psi_f | \hat{H}_{\tmop{IA}} | \Psi_i \rangle$ we need as a second ingredient the molecular wave functions. In the following, we focus on $\rm D _2 \rm S_2$, one of the simplest chiral molecules, which is frequently used in studies of enantioseparation and interconversion \cite{kral03,thanopulos03,trost09,li10b}. It is an almost symmetric prolate top with an asymmetry parameter \cite{zare88}
$\kappa = (2 B - A - C)/(A- C) =  -0.99994 $ 
close to $-1$ ($A =76.15\, \tmop{GHz}$, $B = 6.401\, \tmop{GHz} $, $ C = 6.399\, \tmop{GHz}$ \cite{thanopulos03}). We can thus safely describe its rotation by the Hamiltonian of a symmetric top
\begin{eqnarray}
  & \hat{H}_{\tmop{rot}} & 
    \begin{array}{ll} 
        = & h C \hat{J}^2 +  h\left( A - C \right) \hat{J}_{ z}^2,
  \end{array} 
\end{eqnarray}
where $\hat{J}_{ z}$ is the angular momentum along the symmetry axis of the top and $\hat{J}$ the total angular momentum.
Its eigenstates  $|J K M \rangle $ are determined by the total angular momentum $J$, and by its projections on the molecule-fixed $z$-axis, $- J \leqslant K \leqslant J$, and on the space-fixed $z$-axis $- J  \leqslant M \leqslant J$ respectively.
Using the Euler angles they are given by
\begin{eqnarray}
  \langle \alpha \beta \gamma |J K M \rangle & = & \sqrt{\frac{2 J + 1}{8 \pi^2}} D^{J \ast}_{M K} \left( \alpha, \beta, \gamma \right), 
\end{eqnarray}
where $D^J_{M K} $ are the rotation matrices \cite{zare88}.

Since the coupling between rotations and vibrations can be neglected for the relevant vibrational excitations we take the full wavefunction of the molecule to be a product of the rotation state and the vibrational electronic wavefunction $| v_i \rangle $,
\begin{eqnarray}
  | \Psi_i \rangle & = &  | v_i   \rangle  | J_i K_i M_i \rangle . 
\end{eqnarray}
We can now evaluate the non-zero Rabi frequencies $\Omega_{fi}  =  \langle \Psi_f | \tmmathbf{\hat{\mu}} \cdot \tmmathbf{E}| \Psi_i \rangle$ and discuss the corresponding selection rules
\begin{eqnarray}
\label{eq:rabifrequenz}
\Omega_{fi}   & =  &\langle v_f |   \langle J_f K_f M_f |
                    \tmmathbf{\hat{\mu}} \cdot \tmmathbf{E}| J_i K_i M_i \rangle
                     | v_i \rangle  \\
              & =  &\sum_{\sigma'} \bignone \langle v_f | \mu^M_{\sigma'} | v_i
                    \rangle \sum_{\sigma} \langle J_f K_f M_f | \bignone \bignone D^{1  \ast}_{\sigma \sigma'} | J_i K_i M_i  \rangle E^S_{\sigma} . \nonumber
  \end{eqnarray}
The vibrational matrix elements $\langle v_f | \mu^M_{\sigma'} | v_i \rangle$ can be calculated independently from the rotational matrix elements $\langle J_f K_f M_f  |D^{1 \ast}_{\sigma \sigma'} | J_i K_i M_i  \rangle$; only the component $\sigma'$ of the molecular dipole couples the two. The rotational part of  Eq.~(\ref{eq:rabifrequenz}) yields
\begin{align}
   I & = \langle J_f K_f M_f | \bignone \bignone D^{1 \ast}_{\sigma \sigma'} | J_i K_i M_i \rangle \nonumber\\
   &  = { \int \langle J_f K_f M_f | \Omega \rangle D^{1 \ast}_{\sigma \sigma'} \left(
  \Omega \right) \langle \Omega | J_i K_i M_i \rangle\, \mathrm{d} \Omega}    \label{eq:rotpart}\\
  & = { \int 
  D^{J_f}_{M_f K_f} \left( \Omega \right) D^{1 \ast}_{\sigma \sigma'} \left(
  \Omega \right) D^{J_i \ast}_{M_i K_i} \left( \Omega \right) \mathrm{d} \Omega} \, ,
  \nonumber
\end{align}
where $\Omega = \left( \alpha, \beta, \gamma \right)$ and $\mathrm{d} \Omega = \sin \alpha \, \mathrm{d} \alpha \,  \mathrm{d} \beta\,  \mathrm{d} \gamma$. Using known relations for the $D$-matrices \cite{zare88} we find that the integral is given by a product of Wigner 3j-symbols. 
\begin{eqnarray}
  I = \left( - \right)^{- K_i + M_i + \sigma' - \sigma} \sqrt{\left( 2 J_f +
  1) \left( 2 J_i + 1) \right. \right.} &  &  \nonumber\\
  \times \left(\begin{array}{ccc}
    J_f & 1 & J_i\\
    M_f & - \sigma & - M_i
  \end{array}\right) \left(\begin{array}{ccc}
    J_f & 1 & J_i\\
    K_f & - \sigma' & - K_i
  \end{array}\right)  &  &  \label{eq:selrules}
\end{eqnarray}
The 3j-symbols can be non-zero only if $\Delta J \equiv J_f - J_i = 0, \pm 1$. Moreover, if the molecular dipole is aligned with the molecular $z$-axis ($\sigma' = 0$ in Eq.~(\ref{eq:rabifrequenz})) it follows that $\Delta K = 0$. The selection rule for  $\Delta M \equiv M_f- M_i$  depends on the laser polarisation. For $z$-polarised light $\Delta M = 0$, while $\Delta M = \pm 1$ for circularly polarized light. It will be important in the discussion below that the 3j-symbols appearing in (\ref{eq:selrules}) vanish if there are only zeros in the lower row and $J_i=J_f$,
even when they fulfill the mentioned criteria for allowed transitions.

It is easy to see that one is able to form closed 3-loops with the above selection rules, i.e.\ it is possible to return to the same quantum state $|J K M \rangle$ after three links of allowed electronic dipole transitions. However, due to the small spacing between the rotational levels there are many other non-resonantly driven transitions besides the loop and  it is not clear a priori to what extent they affect the enantioseparation. We will test this numerically in the next section.

\section{Numerical Analysis\label{sec:numerics}}

We proceed to evaluate the time evolution produced by the laser interactions. Since we must account also for non-resonant transitions we cannot use the effective Eq.~(\ref{eq:effschr}), but have to consider the full internal Hamiltonian given in the interaction picture by \cite{kral01}
\begin{eqnarray}
  H_{\tmop{int}} & = & \sum_{A, B}  (\Omega_{A B} \tmmathbf{r})  \mathrm{e}^{- \mathrm{i} \Delta_{A B} t} |A \rangle \langle B| + \text{h.c.} 
\end{eqnarray}
It is determined by the Rabi frequencies discussed above in Eq.~(\ref{eq:rabifrequenz}). The summation runs over the multi-indices $A \equiv (m, J_A, K_A, M_A)$, $B \equiv (n, J_B, K_B, M_B)$ with $m, n$  the vibration states ($m < n$). The quantity $\Delta_{A B} = E_A - E_B + \hbar \omega_{\tmop{AB}}$ is the detuning  of the laser with respect to the levels $A$ and $B$ in the summation $(A<B)$. Unlike for systems without loops, where the detuning can be gauged away, this is not possible for our setup involving loops. 

In the following we do not consider the detailed process of switching on the lasers. Rather we take relevant limiting cases for the initial state: the diabatic and the adiabatic preparation, as well as an important intermediate case.

Depending on the initial state we obtain different expectation values $\langle H_{\rm int} (\boldsymbol{r},t) \rangle$ of the internal Hamiltonian. If this average dipole potential has different spatial dependencies for left- and for right-handed molecules it can be employed for the spatial separation of enantiomers. 

The setup discussed in Ref.~\onlinecite{li10b} consists of three 
laser beams  propagating in the $z$-direction, slightly shifted laterally ($x$-direction) with respect to each other. The optical dipole force acts in the $x$-direction, i.e.\ perpendicular to the lasers, and the resulting dipole force exerted on the molecules is proportional to the time-averaged internal potential. Since the energy peak in the beam center characterizes the strength of the dipole force we can consider the time-averaged value of $\langle H_{\rm int}(t) \rangle$ as a measure of ones ability to perform enantioseparation. In the setup of \textcite{li10b} $\langle H_{\rm int}(t) \rangle$ is approximately  $\Omega_{12}^{\rm max}$, the maximal Rabi frequency at the center of the Gaussian laser beam connecting vibrational levels $| 1 \rangle$ and $| 2 \rangle$, and we will compare the obtained potentials with this value.

\subsection{Adiabatic Preparation \label{sec:adiabatic}}
As a first limiting case, let us assume that the laser fields are switched on adiabatically. Initial eigenstates of the bare internal Hamiltonian thus evolve into eigenstates of the full internal Hamiltonian $H_{\rm int}(t=0)$ including the molecule-laser interactions. 

Unfortunatly, one cannot use this natural choice of states for enantioseparation. This is due to the fact that the spectra of the left- and the right-handed full Hamiltonian are identical. Given an initial thermal population of states one is thus lead to identical potential energies $\langle H_{\rm int} \rangle$. The reason for the isospectrality is that one can always find a unitary transformation $T$ of the rotation state such that the left-handed Hamiltonian is transformed to the right-handed Hamiltonian via $T^\dagger H_{\rm int}^{L} T=H_{\rm int }^{R}$. This transformation $T$ assigns to each handed eigenvector an eigenvector of the opposite handedness and same energy.  

For instance, for light with $\sigma_x,\sigma_y,\sigma_{+1},$ or $\sigma_{-1}$-polarisation one can use the transformation $T | J_i K_i M_i \rangle =(-)^{M_i} | J_i K_i M_i \rangle $. For light with $\sigma_y,$ or $\sigma_x$-polarization the transformation $T | J_i K_i M_i \rangle = (-)^{J_i} | J_i K_i \!-\!M_i \rangle $ will do the job. For setups with different polarisations of the three lasers one can find composed transformations, e.g.\ for the setup considered in Fig.~\ref{fig:finiteTxxz} we find $T | J_i K_i M_i \rangle = (-)^{J_i+M_i} | J_i K_i \!-\!M_i \rangle $.
 
\subsection{Diabatic Preparation}

Next we consider the opposite case that the lasers are turned on very fast. In this diabatic limit the state has no time to change and is still in a chiral eigenstate of the Hamiltonian $H_{\rm rot} +H_{\rm vib}$ without the interaction term. In the presence of a laser field such a state will be subject to different time evolutions for the different enantiomers. However, by utilizing the transformation $T$ from the last section Sec.~\ref{sec:adiabatic} one finds easily that the obtained potential energies $\langle H_{\rm int}^L(t) \rangle =\langle  \Psi^L(t) | H_{\rm int}^L(t) | \Psi^L(t) \rangle $ and  $\langle H_{\rm int}^R(t) \rangle$ are the same if $\Psi^L(t=0)=\Psi^R(t=0)$. Therefore a diabatic choice of the initial state is not useful for enantioseparation as well.

\begin{figure*}[t,b,p]
\includegraphics[width=0.98\textwidth]{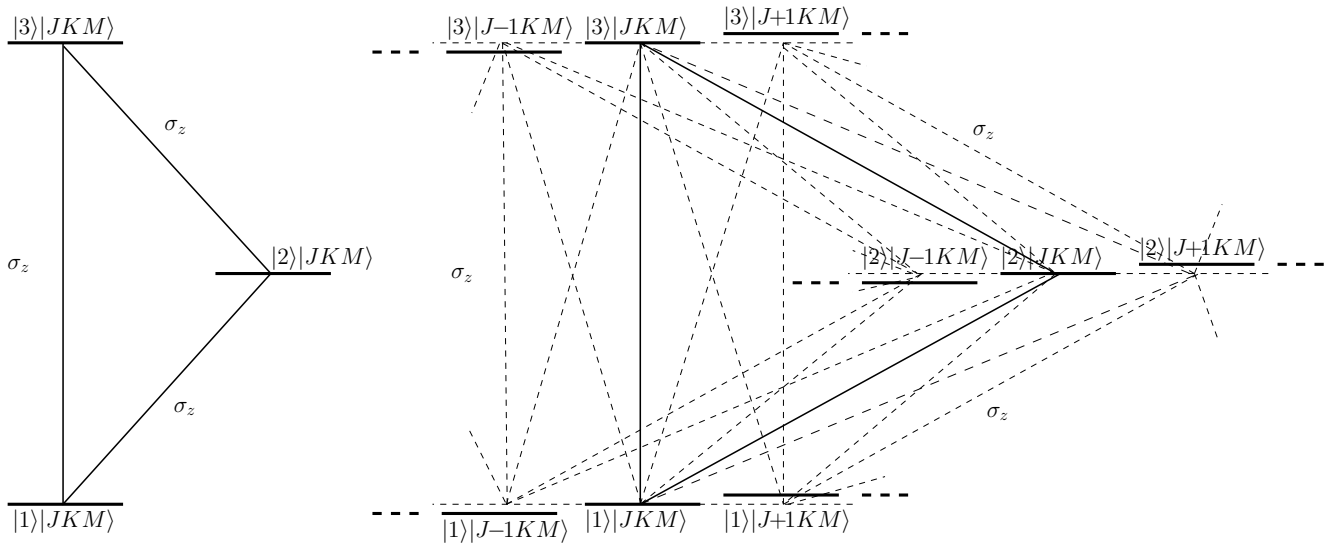}
    \caption{Left: Example of a 3-loop with rotational substates. It is obtained if all lasers are $z$-polarised. Right: The dashed lines show the numerous other transitions necessarily involved when the rotational sublevels are included. The solid line indicates the loop from the left side. Here the molecular dipole is assumed to be aligned along the symmetry axis of the molecule ($z$-direction) which restricts the number of allowed transitions. \label{fig:finiteTzbothfig}}
\end{figure*}

\subsection{Preparation of partially dressed states\label{sec:partdress}}

In this section we discuss a third class of initial states that lies in between the ones just mentioned: vibrationally dressed states with uncorrelated rotations. This is the closest generalization of the dressed states considered in \textcite{li10b}. The vibrational states $| D \rangle$ are eigenstates of $H_{\rm vib} + H_{\rm IA}$, whereas the rotations are in a thermal state, such that 
$ \rho_{\rm tot} = | D \rangle \langle  D | \otimes \rho_{\rm rot}^{\rm therm} $.

Now we consider the possible polarisation configurations. We take the molecular dipole to be  aligned along the $z$-axis of the molecular coordinate system. This implies $\Delta K =0$, such that about one third of all transitions involved in Eq.~(\ref{eq:rabifrequenz}) do not contribute. The simplest loop structure can be found using the $z$-polarisation for all three lasers. As shown in Fig.~\ref{fig:finiteTzbothfig} (left), the loop then consists of levels with the same rotation state (there are few exceptions where this is not possible, e.g.\ for $|J K M \rangle =  | J 0 0 \rangle $). However, at the same time there are many more allowed transitions between the rotational states, see Fig.~\ref{fig:finiteTzbothfig} (right).

\begin{figure}[t,b,p]
  \begin{center}
    \includegraphics[width=0.47\textwidth]{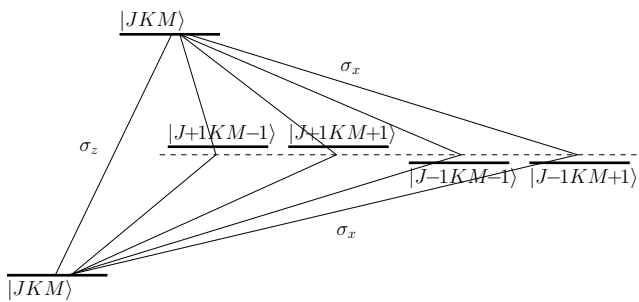}
  \end{center}
  \caption{For most rotational states (except for a few ones such as $|JKM \rangle = |000 \rangle$) it is possible to form several closed loops, e.g., with two  $x$-polarised and one $z$-polarised laser beams\label{fig:finiteTxxz}}
\end{figure}

We note that, in the special setup chosen by Ref.~\onlinecite{li10b} the three laser beams cannot be polarised in the $z$-direction, since this is their propagation direction. Choosing alternative non-$z$-polarisations is not an option. It is not possible to form a closed 3-loop by using just lasers of $x$, $y$, $\sigma_+$ or $\sigma_-$ polarisation, since the selection rule for $M$ is then $\Delta M = \pm 1$, which cannot lead to the original state. However, using at least one laser with $z$-polarisation ($\Delta M = 0$) we can obtain again closed loops as shown in Fig.~\ref{fig:finiteTxxz}. We will use the later setup in our numerical simulations, noting that similar results are obtained for other choices of the polarisations.

\textit{Time scales.} The time scales involved in the setup are (a) the time $\tau_\Omega$ associated with the Rabi frequency $\Omega_{12}^{\rm max}$ describing the vibrational population transfer ($\approx 4.8 \,\rm ns$ for $\Omega_{12}^{\rm max}= \sqrt{Q}\cdot 1\cdot 10^{-9} {\,\rm hartree} $ with $Q=1000$ as in Ref.~\onlinecite{li10b}), (b) the time scale $\tau_\Delta$ of the rotational constants (for $\rm D_2 S_2$:   $1 /A=0.013\,\rm ns$ and $1 /B\approx 1 /C=0.154\,\rm ns$), (c) $\tau_{\rm exp}$, the time scale of the whole experiment (typically $10-40\, \rm \mu s$), and (d)  the tunneling time $\tau_{\rm LR}$ of the $\rm D_2S_2$ molecule ($33\,\rm ms$). We have a clear separation of time scales $\tau_\Delta < \tau_\Omega \ll \tau_{\rm exp} \ll \tau_{\rm LR}$.

\begin{figure}[t,b,p]
   \includegraphics[width=0.47\textwidth]{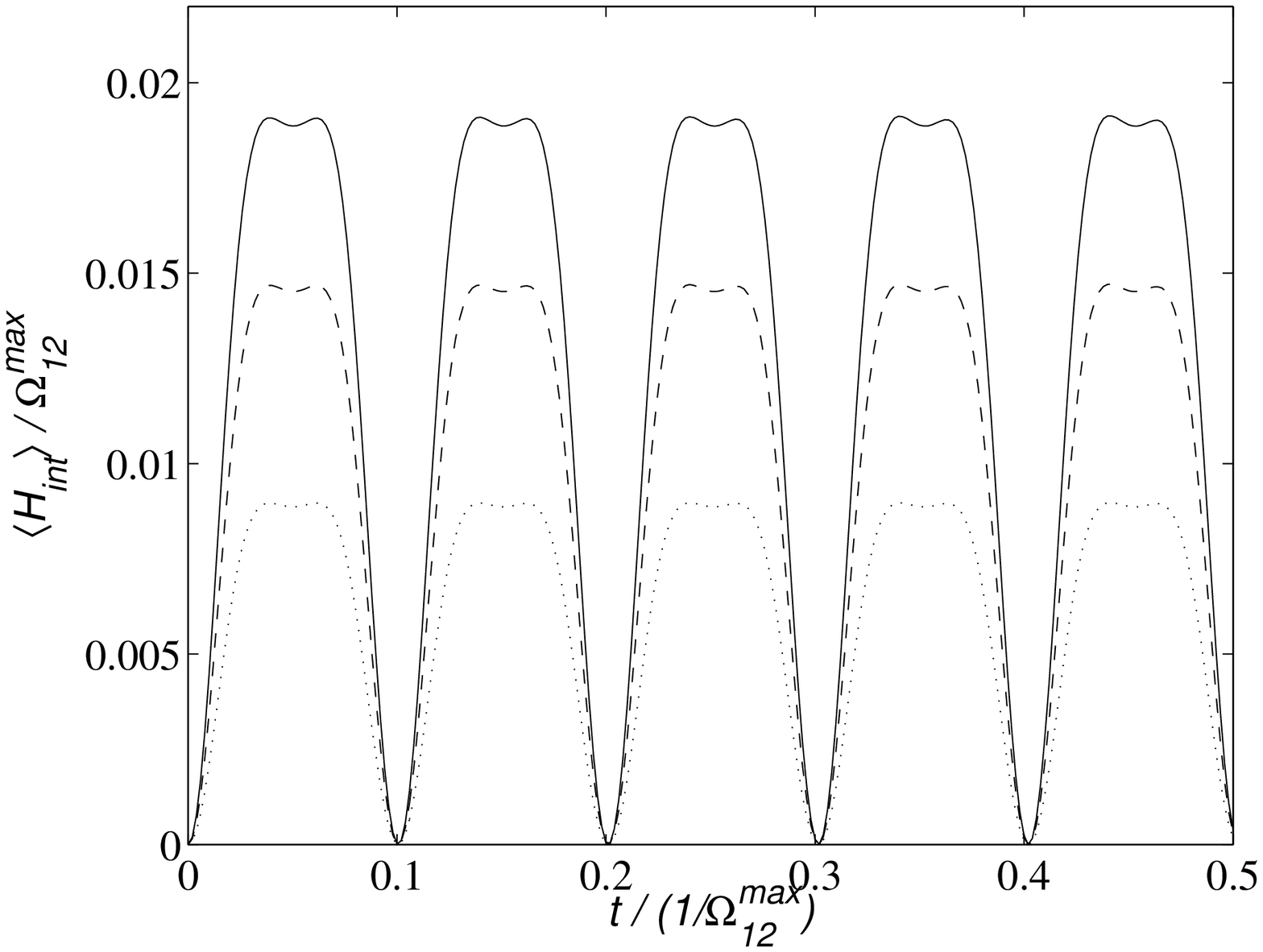}  
   \includegraphics[width=0.47\textwidth]{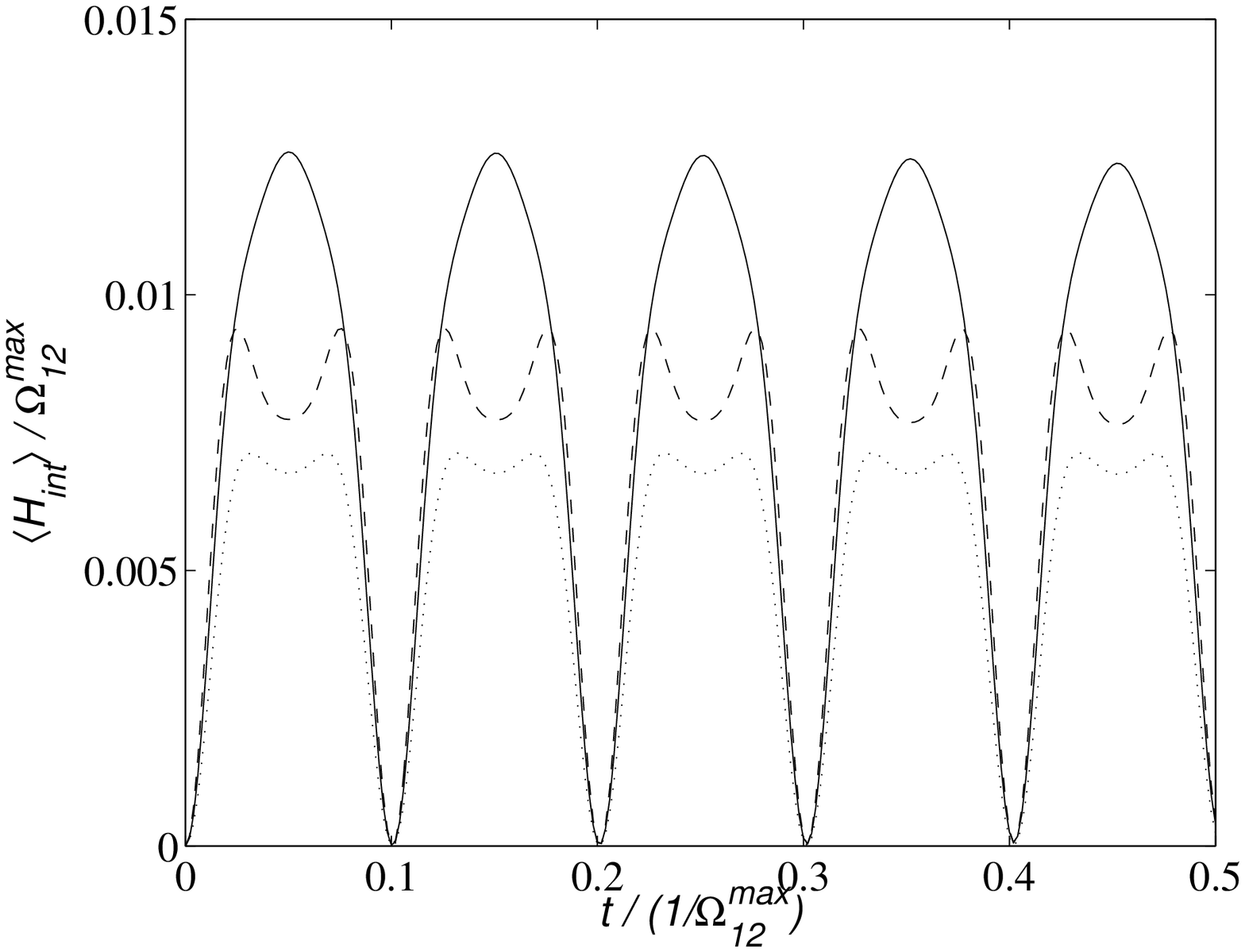}
\caption{The interaction energy for right- and left-handed molecules oscillates due to the time-dependent Hamiltonian. The solid line depicts the effective potential if the system is initially in the dressed states $| \chi_1^R \rangle$ and $| \chi_2^L \rangle$, the dashed lined is for $| \chi_2^R \rangle$ and $| \chi_1^L \rangle$, the dotted line for $| \chi_3^R \rangle$ and $| \chi_3^L \rangle$. Upper figure: the lasers are resonant with respect to the rotational ground state of each vibrational state. Lower figure: the lasers are resonant with respect to the transitions $|1 \rangle | J K M \rangle \leftrightarrow |2 \rangle |  J\!\!+\!\!1 K M \rangle \leftrightarrow |3 \rangle | J K M \rangle$. Parameters: $T=0.5\, \rm K$, lasers polarised as in Fig.~\ref{fig:finiteTxxz} ($\Omega_{13}$ $z$-polarised, $\Omega_{12}$ and $\Omega_{23}$ $x$-polarised).   \label{fig:Tfinitexxzboth}}
\end{figure}

\textit{Numerical observations.} The time scales just discussed can be seen in our numerical simulations. In Fig.~\ref{fig:Tfinitexxzboth} we plot the potential energy for the center-of-mass motion in units of the Rabi frequency $\Omega_{12}^{\rm max}$ versus the time in units of $1 /\Omega_{12}^{\rm max}$. One observes that the energy oscillates at the scale of the involved detunings $\Delta = B \left( J_f^2 - J_i^2 + J_f - J_i \right) + \left( A - B \right) \left( K_f^2 - K_i^2 \right)$, typically one order of magnitude faster than the time scale associated with $\Omega_{12}^{\rm max}$. It is difficult to identify each frequency in detail due to the complex behaviour of the dynamics in a closed loop network compared to a 2-level system.

The potentials experienced by right- and left-handed molecules now differ, but we observe that the strength of the potential is reduced by about to two orders of magnitude as compared to the case where the rotations are neglected. After the time average this remaining potential seems to be way too small for the suggested separation scheme of \textcite{li10b}, see Fig.~\ref{fig:Tfinitexxzboth}. The observed potentials are characteristic for similar setups. 

So far the lasers are taken to be resonant with respect to the same rotational substates. Even if we change the laser frequency slightly to make one of the loops resonant, the oscillations caused by the other transitions show a similar behaviour, see lower part of Fig.~\ref{fig:Tfinitexxzboth}. Likewise, a different orientation of the  dipole in the molecular  frame will not influence the result. 

We find that the chiral sensitivity survives, unlike in the adiabatic and diabatic case above, even though the partially dressed state yields highly oscillating potentials. Whether this remaining effect could lead to a feasible experimental setup is open to future studies.

\begin{figure}[t,b,p]  
   \includegraphics[width=0.47\textwidth]{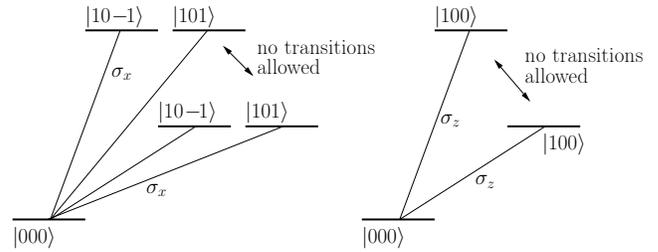}
  \caption{For the zero temperature case we start our simulation in the rotational ground state {$|JKM \rangle = |000 \rangle$}. Due to the selection rules it is not possible to form a closed loop in the rotational state manifold using any combination of polarisations. The left diagram shows the allowed transitions for $x$-polarisation, the right one for $z$-polarisation.\label{fig:T0xxandzz}}
\end{figure}

\begin{figure}[t,b,p]
   \includegraphics[width=0.47\textwidth]{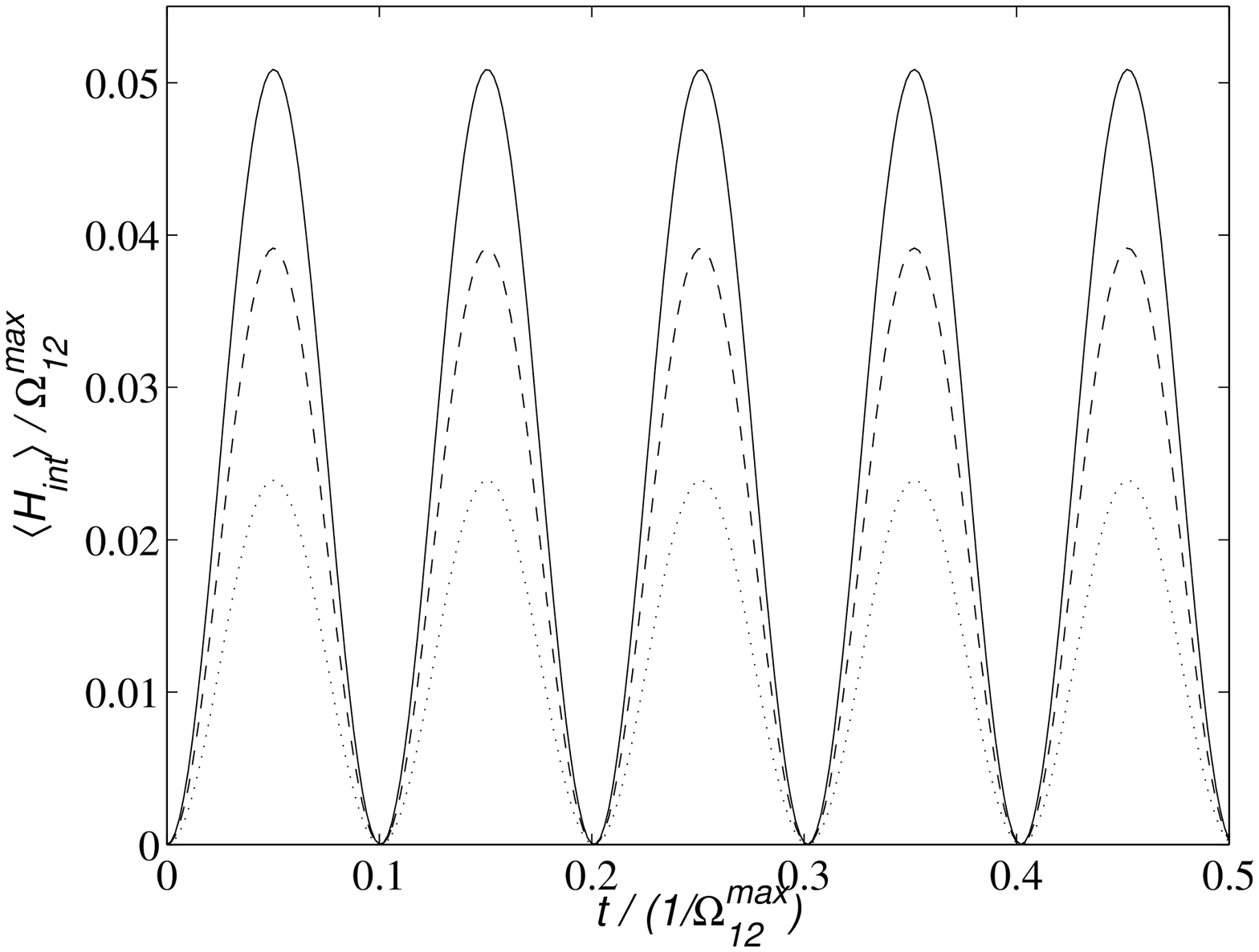}
   \includegraphics[width=0.47\textwidth]{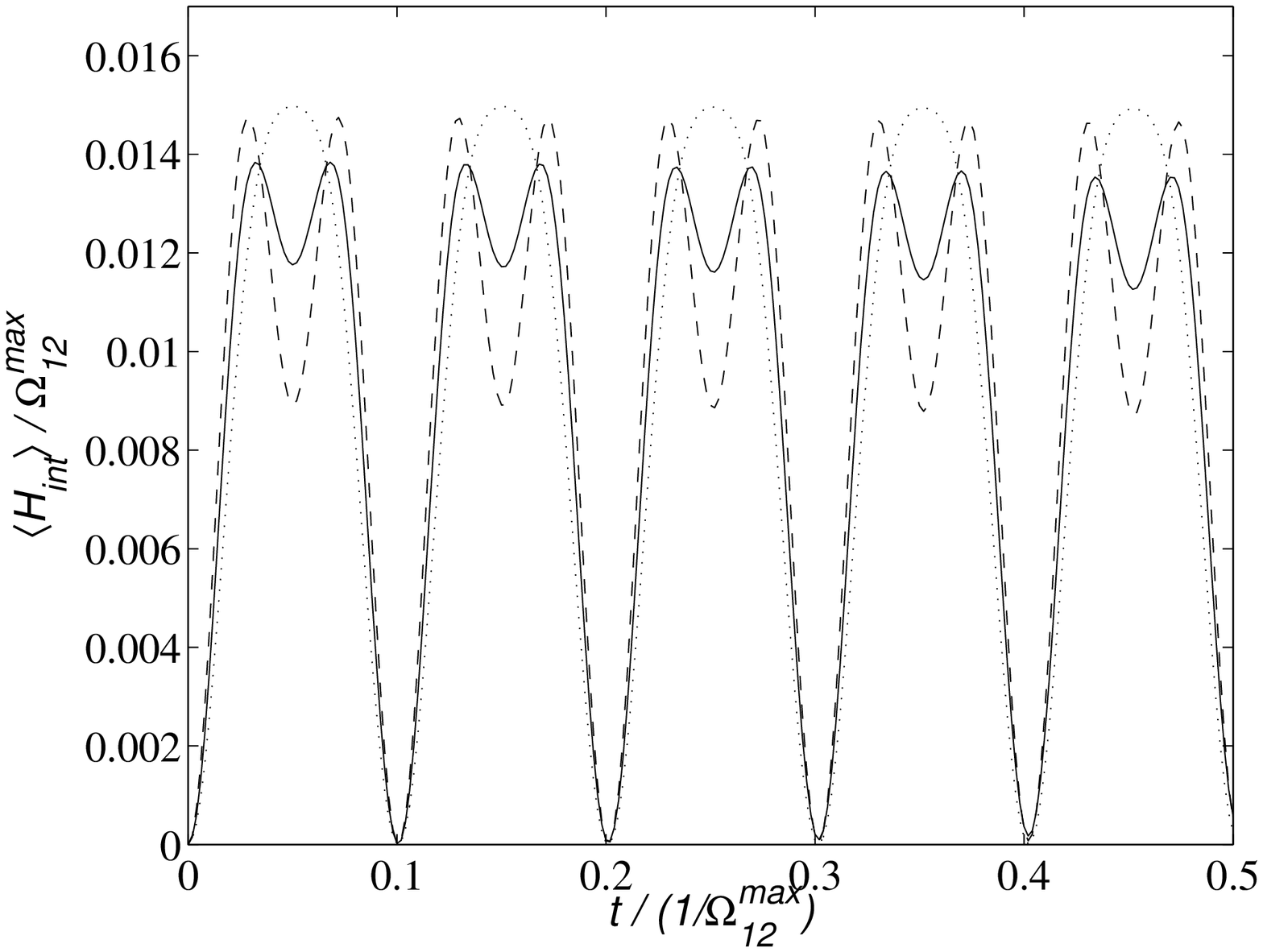}
 \caption{The interaction energy for right- and left-handed molecules oscillates due to the time-dependent Hamiltonian. The solid line depicts the potential if the system is initially in the vibrationally dressed states $| \chi_1^R \rangle$ and $| \chi_2^L \rangle$, the dashed lined is for $| \chi_2^R \rangle$ and $| \chi_1^L \rangle$, the dotted line for $| \chi_3^R \rangle$ and $| \chi_3^L \rangle$, see Sec.~\ref{sec:std}. Upper figure: the lasers are resonant with respect to the rotation ground state of each vibrational state. Lower figure: the lasers are resonant with respect to the transitions $|1 \rangle | J K M \rangle \leftrightarrow |2 \rangle | J\!\!+\!\!1 K M \rangle \leftrightarrow |3 \rangle | J K M \rangle$. Parameters: $T=1\,\rm m K$, lasers polarised as in Fig.~\ref{fig:finiteTxxz} ($\Omega_{13}$ $z$-pol., $\Omega_{12}$ and $\Omega_{23}$ $x$-pol.). 
  \label{fig:1mKxxzboth}}
\end{figure}

\textit{Zero Temperature Case.} Next we briefly discuss the special case of $T=0$, where the system is in the rotational groundstate $| JKM \rangle = |000 \rangle$. This corresponds approximately to an assumed temperature of $1\,\rm mK$ \cite{li10b}. However, the selection rules Eq.~(\ref{eq:selrules}) forbid the transition $|JKM \rangle \to |JKM \rangle $ for $z$-polarized light if $|JKM \rangle = |000 \rangle  $. Therefore, one cannot form a closed 3-loop starting from $|000 \rangle$ with allowed transitions, see Fig.~\ref{fig:T0xxandzz}. Irrespectively of that we can have a  look at this case as well. In order to facilitate the comparison with previous results we use in our numerical simulations  the same laser configuration as for the finite temperature case in Fig.~\ref{fig:finiteTxxz}. 
The obtained potentials reported in Fig.~\ref{fig:1mKxxzboth} are qualitatively similar to the finite temperature case of Fig.~\ref{fig:Tfinitexxzboth}. That is, the potentials oscillate at a time scale of the detunings yielding averaged potentials of about two orders of magnitudes below the $\Omega^{\rm max}_{12} $.

This result suggests that it is possible to obtain chiral sensitive potentials without having a 3-loop in the Hamiltonian. On the other hand, it seems difficult, if not practically impossible, to create the initial dressed state in the vibrational manifold without having a closed loop.

\textit{Time-independent  potentials.} The time-independent  potentials of \textcite{li10b} can be recovered up to the factor Eq.~(\ref{eq:rotpart}) if we restrict the system artificially to only one 3-loop with no connections to other states. The maxima of these potentials are only reduced by the orientation factor Eq.~(\ref{eq:selrules}) in the selection rules, which is typically between $\pm 0.1$ and $\pm 0.5$. This way consistency with proposal in Ref.~\onlinecite{li10b} is obtained at the expense of treating an unphysical situation.

\section{Conclusions}
In this paper we highlighted the importance of the rotational state on enantioseparation. The studied enantioseparation scheme is based on the sign difference of the Rabi frequencies of two enantiomers. As compared to previous studies \cite{li10b}, we consider a more realistic molecular description, including the orientation state.
We find that the ability to create chirality-dependent potentials depends strongly on the preparation of the initial states. For a usual adiabatic and diabatic preparation we find no chiral dependence, whereas for a partially dressed state chiral dependence can be found. 
 However, due to the detunings, the time-dependent Hamiltonian leads to a {\itshape time-dependent} potential. The oscillations occur at the time scale of the molecular rotations. 
We observe that even in the rotational ground state the time-average of the resulting potentials is typically two orders of magnitude smaller  than the potentials for a molecule with fixed orientation, which is the relevant quantity for enantioseparation. 

We thank Hendrik Ulbricht for discussions. The authors acknowledge support by the MIME project within the ESF Eurocore EuroQUASAR program.

\end{document}